# Dynamic spending and portfolio decisions with a soft social norm

Knut Anton Mork*, Fabian Andsem Harang**, Haakon Andreas Trønnes***, and Vegard Skonseng Bjerketvedt****

November 2022

* Professor Emeritus, Norwegian University of Science and Technology (NTNU), Department of Economics, knut.anton.mork@ntnu.no

** Professor, BI Norwegian Business School, Department of Economics, fabian.a.harang@bi.no

**PhD student, Norwegian University of Science and Technology (NTNU), Department of Economics, haakon.a.tronnes@ntnu.no

***PhD student, Norwegian University of Science and Technology (NTNU), Department of Industrial Economics and Technology Management, vsb@ntnu.no

Withdrawal smoothing
Dynamic risk taking
Fund sustainability

We are indebted to three referees for exceptionally helpful comments and recommendations. Thanks to Snorre Lindset for encouraging us to pursue this topic and to Alfonso Irrarazabal for guidance regarding numerical solution methods. Espen Henriksen has offered useful comments. The research has benefitted from financial assistance from Finansmarkedsfondet, the financial markets research fund under The Research Council of Norway, Project no. 294398.

ABSTRACT

We explore the implications of a preference ordering for an investor-consumer with a strong preference for keeping consumption above an exogenous social norm, but who is willing to tolerate occasional dips below it. We do this by splicing two CRRA preference orderings, one with high curvature below the norm and the other with low curvature at or above it. We find this formulation appealing for many endowment funds and sovereign wealth funds, including the Norwegian Government Pension Fund Global, which inspired our research. We solve this model analytically as well as numerically and find that annual spending should not only be significantly lower than the expected financial return, but mostly also procyclical. In particular, financial losses should, as a rule, be followed by larger than proportional spending cuts, except when some smoothing is needed to keep spending from falling too far below the social norm. Yet, at very low wealth levels, spending should be kept particularly low in order to build sufficient wealth to raise consumption above the social norm. Financial risk taking should also be modest and procyclical, so that the investor sometimes may want to "buy at the top" and "sell at the bottom." Many of these features are shared by habit-formation models and other models with some lower bound for consumption. However, our specification is more flexible and thus more easily adaptable to actual fund management. The nonlinearity of the policy functions may present challenges regarding delegation to professional managers. However, simpler rules of thumb with constant or slowly moving equity share and consumption-wealth ratio can reach almost the same expected discounted utility. However, the constant levels will then look very different from the implications of expected CRRA utility or Epstein-Zin preferences in that consumption is much lower.

# 1. Introduction

This paper explores the normative implications for a long-term investor whose preferences are influenced by an exogenous social norm. If the norm is broken, the curvature of the utility function increases discontinuously so that further declines in consumption are experienced as progressively painful, though finitely so. The role of the norm resembles that of habit in habit-formation models but is softer in the sense that utility does not immediately fall to negative infinity upon breaking the norm. We believe such a formulation can serve as a reasonable approximation to the preferences of owners of many endowment funds and sovereign wealth funds, including the Norwegian Government Pension Fund Global (GPFG), which inspired our work. This fund, with a current worth of about USD 1.3 trillion, serves as a major source for the annual government budget that seeks to maintain government services on levels comparable to the norms implied by voters' expectations.

Our analysis finds that a consumer-investor with the preferences we specify will be significantly more risk averse than one unconstrained by a social norm. We also find that the optimal risk taking will be procyclical. Both features resemble the corresponding ones of habit-formation models but are less extreme. Furthermore, whereas the habit-formation models severely constrain consumption in good times in order to ensure that the habit is satisfied in bad times, our specification is less restrictive in good times and tells, we believe, a credible story of how utility suffers when the norm is broken in bad times. Perhaps the most intriguing implication of our specification is a very high propensity to save at very low wealth levels because the incentive in terms of potential future utility gains outweigh the short-term sacrifice.

The issues facing long-term institutional investors have received considerable attention in the recent literature, with Campbell and Sigalov (2022) and Cochrane (2022) as leading examples. On a basic level, optimal plans for spending and portfolio over time for an infinite-horizon investor were derived by Merton (1971). His solution was appealing in its simplicity, namely, to spend a constant share of wealth and maintain a constant risk profile (equity share) regardless of what happens in the financial markets. These conclusions follow readily from his assumption of power (or log) utility.

However, the subsequent literature has pointed out a number of additional issues that are or should be relevant for agents concerned about long-term developments. Cochrane (op.cit.) advocates a shift in focus from current asset values to payoffs in the form of coupons, dividends, and share buybacks. Others have focused on sustainable consumption, such as Arrow et al. (2004), who, in response to the Brundtland report (World Commission on Environment and Development, 1987), presented and analyzed criteria for sustainable consumption on the global scale. Campbell and Martin (2022) recently added to this line of research by specifying sustainability as a constraint on investment and consumption decisions, requiring that welfare should not be expected to decline over time. This constraint does not distort portfolio choice but does impose an upper bound on the consumption-wealth ratio as well as the rate of time preference.

A more commonly used criterion for sustainability, codified, for example for the GPFG, is to limit spending from a fund to the fund's expected real financial return. As is well known, such a rule keeps the expected future fund value equal to its current value; however, as Dybvig and Qin (2021) have shown, the fund will nevertheless be depleted almost surely because the probability mass for the future fund size will ultimately collapse into a spike next to zero[1].

Although this problem can be overcome by calibrating withdrawals to the geometric rather than the arithmetic rate of return, the above rule also has the disadvantage that it occasionally can call for sharp spending cuts. For example, a 20 percent drop in asset prices translates into a 20 percent spending cut. To avoid such choppiness, many institutions allow smoothing, for example in the form of the so-called MIT-Tobin rule, which stipulates spending as a weighted average of the expected return and last year's withdrawal, often with a weight of as much as 0.8 for the latter[2]. For Norway's GPFG, the Fiscal Rule of 2001 contains a less specific proviso allowing for smoothing of withdrawals for cases in which the main rule of withdrawing an amount equal to the expected financial return implies sharp changes in withdrawals from one year to the next[3].

This problem has led various authors to propose alternative smoothing mechanisms. The literature on habit formation builds penalties into the utility function in the form of fast rising marginal utility if spending is cut relative to an established habit. Habits can be formed internally as an aggregation of the agents' own consumption (Abel, 1990; Constantinides, 1990), or as an attempt to keep up with habits formed in society at large (Abel, op.cit.; Campbell and Cochrane, 1999). Such models have been used in the macroeconomics literature to account for the so-called excess smoothness of consumption (e.g. Campbell and Deaton, 1989 and Galí, 1990).

---

[1] Mork and Trønnes (2022) show that the future fund value will not even be preserved in expectation if the financial returns are negatively serially correlated. This is likely to be true for a fund domiciled in a small country that is invested in the global market in foreign currencies with spending withdrawals based on the fund's value in the domestic currency if the exchange rate obeys long-term purchasing parity. Mork, Trønnes, and Bjerketvedt (2022) find this effect to be quantitatively important for the GPFG.
[2] http://web.mit.edu/fnl/volume/205/alexander_herring.html, accessed on Dec 1, 2020. See also Tobin (1974).
[3] https://www.regjeringen.no/no/dokumenter/stmeld-nr-29-2000-2001-/id194346/, accessed October 5, 2022. It is worth noting that the Act governing the GPFG does not allow temporary borrowing to be used to smooth spending, so that all smoothing is applied directly to the withdrawals, https://www.regjeringen.no/contentassets/9d68c55c272c41e99f0bf45d24397d8c/government-pension-fund-act-01.01.2020.pdf, accessed October 5, 2022.

In Constantinides' (1990) model, the optimal risky portfolio share is an increasing function of the riskless rate. The same is true of the model in Dybvig (1995, 1999), which starts with Merton's (1971) model, but adds a constraint requiring that consumption never declines over time, echoing Duisenberry's (1949) idea about household behavior. In both models, the riskless part of the portfolio needs to be large enough to secure a safe stream of revenue to fund the habit (in Constantinides's case) or keep consumption from falling (Dybvig's). Because a higher riskless rate eases this constraint, this feature makes risk taking a non-decreasing function of the riskless rate. Models with drawdown constraints on consumption, such as those of Shin et al. (2007), Arun (2012), and Jeon and Park (2021), have similar implications.

Empirically, Campbell and Sigalov (2022) argue that the long decline in risk-free rates in recent decades have made a number of large endowment funds and sovereign wealth funds (including the GPFG) move in the opposite direction. They seek to explain this behavior by presenting a model that starts with Morton's (1971) specification but replaces the optimization of consumption with a constraint that consumption be sustainable in the sense of matching the expected rate of financial return. They then find that declining interest rates make such agents reach for yield by taking more risk.

The habit-formation models, drawdown-constrained models, and Dybvig's model of non-decreasing consumption carry the important implication that portfolio risk taking be procyclical. The reason is that a financial loss tightens the constraint that a sufficient part of the portfolio must be risk free so as to secure funding of the habit or non-decreasing consumption, respectively. Thus, fund owners that care about the smoothness of spending should be cautious about risk taking, especially in bad times, while constraining consumption in good times.

Instructive as these models are, they carry an important weakness by placing a hard lower bound on the acceptable level of spending. If the required riskless set-aside exceeds the value of the entire portfolio, utility drops abruptly to minus infinity, which essentially means that the model collapses. Although the constraint on good-times consumption prevents this from happening in the model, the hard lower bound on spending becomes a very real issue for real-world fund owners who might consider transitioning to the rules implied by a habit-formation or similar model. Suppose, for example, that the Norwegian GPFG, whose current set of rules are similar to the sustainability principle in Campbell and Sigalov (2022), were to switch to rules corresponding to a model such as Constantinides (1990) or Dybvig (1995). The fund, whose value at this writing is about USD 1.3 trillion, contributed USD 30 billion to the 2022 government budget. To prevent this contribution from declining, as in Dybvig's case, the entire fund would have to be invested without risk if the real riskless rate is 2.5 percent—a rather generous estimate, at least until recently. With Constantinides' model, the habit level would likely be lower than the actual spending, so that the riskless share could be somewhat less than 100 percent. Even so, the implication for risk taking would contrast sharply with the fund's current mandate, according to which 70 percent of the fund can be invested in equity or real estate and another 9 percent in corporate bonds.

Our model sidesteps this problem. Rather than imposing a hard constraint of non-decreasing consumption or an equally hard constraint to keep spending from falling below a certain level, we propose a specification where the marginal utility of consumption rises progressively more sharply if consumption falls below a certain point. Although the agent will experience such a decline as especially painful, it does not make utility fall to minus infinity. Because we specify this limit as an exogenously given time trend, we do not refer to it as habit (although in a previous version we did), but as a social spending norm. Because consumption below this norm is merely painful and not literally unbearable, we refer to it as a soft social norm.

Technically, we do this by splicing two power (CRRA) utility functions that are identical except for the value of the curvature parameter $\gamma$. If consumption falls below the social norm, this parameter jumps to a higher value. This feature makes risk taking procyclical and lower than if the lower $\gamma$ value were valid everywhere, yet higher than in habit-formation and similar models. Consumption is also higher in good times. A rather unique feature is a willingness to make large sacrifices of consumption at very low wealth levels so as to build sufficient future wealth to keep the probability low that future consumption falls below the future social norm.

In a wider context, our analysis makes a contribution to the literature on sovereign wealth funds and institutional endowment funds. Among recent studies, Braunstein (2022) discusses how the purpose of the various funds differ depending on the political context. Earlier surveys have been written by Baldwin (2012) and Alhashel (2015). Arouri et al. (2018) similarly discusses their varying investment strategies, as did Bernstein et al. (2013), Paltrinieri and Pichler (2013), Dreassi et al. (2017) and Johan et al. (2013). Empricial studies of endowment funds include Barber and Wang (2013) Brown, et al. (2014), and Dahiya & Yermack (2018). A special issue has been taken up by van der Bremer et al. (2016). and Irarrazabal et al. (2022), who discuss the joint decision of exhaustible-resource extraction and portfolio choice for a sovereign wealth fund established to safeguard the revenues for that extraction for future generations. We bypass that issue and study instead the joint decision of portfolio management and revenue spending for an already established fund without consideration of other possible non-tradeable assets.

The next section presents the utility function that we use to represent the soft social norm and explores its analytical properties. In addition to setting up the model and exploring some key features of the solution, it also outlines an explicit analytical solution, for which more detail is offered in the Appendix. Although this solution offers some important intuitive insights, its complexity makes numerical methods preferable for exploring the model's main implications. Section 3 derives, presents, and analyzes such a numerical solution. Section 4 discusses the result, partly by comparing them to those of habit-formation and other similar models, partly by highlighting the possible implications for sovereign wealth funds, and partly by considering how the actions suggested by our model can be delegated and carried out in practice. Section 5 summarizes the results and concludes.

## 2. The Model

Our idea of a soft social norm involves a certain norm of the consumption level, denoted by $x > 0$, and a utility function whose curvature becomes flatter when consumption increase above this norm. A natural starting point might be a specification of marginal utility of the form

$$u'(c) = c^{-\gamma(c)},$$

where $\gamma$ is a decreasing function, possibly depending on the norm $x$. If this function is differentiable, the standard curvature measure becomes

$$-\frac{u''(c)c}{u'(c)} = \gamma(c) + \gamma'(c)c \ln c.$$

Note however that this measure might be negative for certain levels of consumption, depending on the structure of the chosen $\gamma$, an undesirable feature for our modelling purposes.

To overcome this issue while still preserving some simplicity of the model we therefore propose to choose the risk-aversion function $\gamma$ as a step function, falling discontinuously from a higher to a lower value when consumption falls below a certain level we identify as the social norm. That is, we start by considering two CRRA utility functions, differing only in the attitude towards risk:

$$u_j(c) = \frac{c^{1-\gamma_j}}{1-\gamma_j}, \quad j = 1,2; \quad \gamma_1 > \gamma_2. \tag{1}$$

We refer to these preference orderings as CRRA$_1$ and CRRA$_2$, respectively. We assume that the agent has access to two assets, one risky asset that we refer to as equity and that is described by a geometric Brownian motion with mean $\mu$ and volatility $\sigma$, and one riskless asset with constant return $r$. These two assets are combined by investing a share $\omega$ in the risky asset and $1 - \omega$ in the riskless asset. The portfolio constructed by the investor is self-financing. The investor is allowed to consume at rate $c$ from total wealth, and thus the wealth of the investor is described by the stochastic differential equation

$$dW_t = W_t\big(r + \omega_t(\mu - r)\big)dt + W_t \sigma dB_t - c_t, \quad W_0 = w.$$

Here $\{B_t\}_{t\geq 0}$ describes Brownian motion. Given a utility function $u$, the objective of the agent is to maximize expected utility of discounted consumption over an infinite time horizon, i.e. to maximize the following payoff function over any admissible $c$ and $\omega$

$$P(t,w|c,\omega) = \mathbb{E}\left[\int_t^\infty e^{-\rho(s-t)} u(c_s) ds | W_t = w\right].$$

If we choose the utility function $u = u_j$ for $j = 1,2$, the optimal equity share would be, as shown by Merton (1971),

$$\omega_j^* = \frac{\pi}{\gamma_j \sigma^2}, \qquad (2)$$

where $\pi = \mu - r$ is the equity premium.

Optimal rate of consumption would similarly be given by the Keynes-Ramsey rule:

$$c_j^*/w = \eta_j \equiv (1/\gamma_j)\rho + (1 - 1/\gamma_j)\left(r + \frac{1}{2}\omega_j^*\pi\right), \qquad j = 1,2, \qquad (3)$$

where $\rho$ is the subjective discount rate, and $w$ represents the value of the portfolio.

### 2.1. Spliced utility function

We now use the two preference orderings in (1) as building blocks in the specification of the preferences of our consumer-investor agent. That is, we splice these orderings such that the curvature jumps discontinuously at the point where consumption equals the soft social norm $x$:

$$u(c,x) = \begin{cases} \frac{1}{1-\gamma_1}\left[\left(\frac{c}{x}\right)^{1-\gamma_1} - 1\right], & c < x \\ \frac{1}{1-\gamma_2}\left[\left(\frac{c}{x}\right)^{1-\gamma_2} - 1\right], & c \geq x. \end{cases} \qquad (4)$$

This function is continuous and concave everywhere in its first argument. It is increasing in consumption and decreasing in the social norm and the mapping $\alpha \mapsto u(\alpha c, \alpha x)$ can easily be seen to be homogeneous of degree zero.

Let $u_c$ denote the derivative in the first variable of the function $u$ and $u_x$ the derivative in the second variable. One can readily check that $u$ is such that the first partial derivative with respect to $c$ satisfies $u_c(c,x) = c^{-\gamma(c,x)}$, where

$$\gamma(c,x) = [\gamma_1 + (1-\gamma_1)log_c(x)]\mathbb{1}_{c<x} + [\gamma_2 + (1-\gamma_2)log_c(x)]\mathbb{1}_{c\geq x}.$$

Furthermore, the first derivative in the first variable of $u$ is continuous on the positive real numbers and continuously differentiable everywhere. At the point $c = x$ the first-order derivative has a kink. The second-order derivatives are continuous everywhere except in the point $c = x$, where a jump happens. The main attractive feature of this utility function is that at $c = x$, the standard curvature measure jumps from $\gamma_1$ to $\gamma_2$. It may also be worth noting that utility has the same sign as $c - x$.

*Figure 1: Marginal utility with a soft social norm*

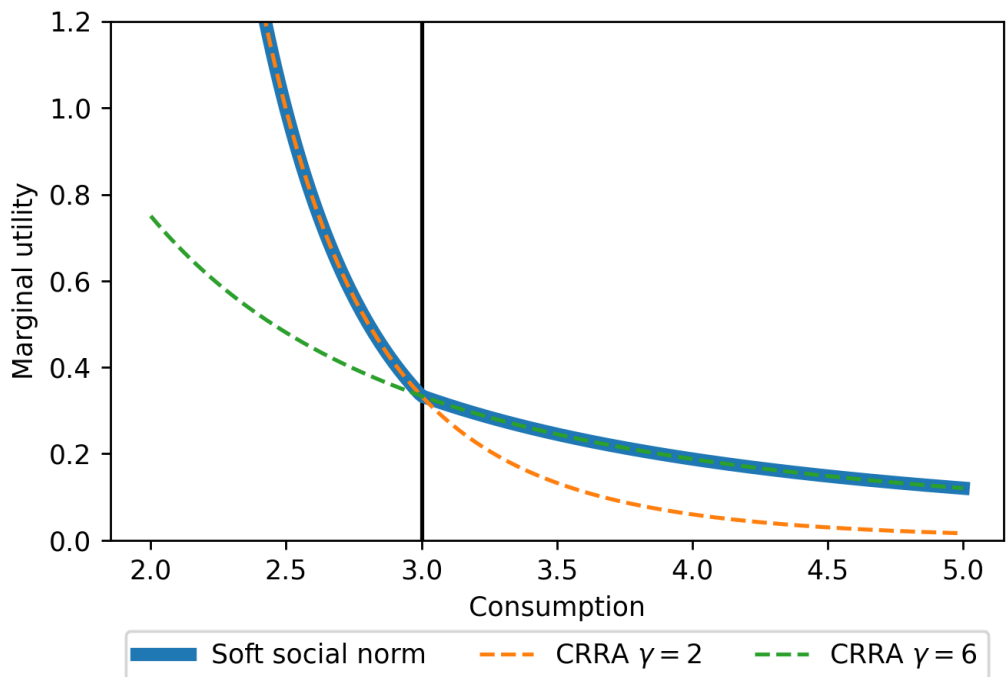

Figure 1 displays the marginal utility of an agent with a soft social norm of $x = 3$, and of the two associated CRRA agents. The graph is drawn on the assumption that $\gamma_1 = 6$ and $\gamma_2 = 2$.

Thus, this agent is extremely averse to consumption $c$ falling below $x = 3$. Yet, the agent can accept such an outcome if the alternatives are worse.

For comparison, we note that Constantinides' (op. cit.) habit utility function can be written in our notation as

$$u^H(c,x) = \frac{(c-x)^{1-\gamma}}{1-\gamma}, \qquad (5)$$

where the function is only defined for $c \geq x$. Although this is not a special case of our specification, it is somewhat similar in that both specifications imply sharply larger marginal utility for low levels of consumption.

### 2.2. Dynamic setup

For a fixed habit level $x$, let the value function $V$ be described by the optimization $V(t,w,x) = \max_{c,\omega} P(t,w|c,\omega)$, where the payoff function $P$ is built with the utility function defined in (4). Because our model is partial, we assume that the social norm is dynamic, exogenously given, and that it moves with a constant drift $g$, so that

$$dx_t = x_t g \, dt, \quad x_0 = x. \qquad (6)$$

In continuous time, the associated Hamilton-Jacobi-Bellman (HJB) equation describing the dynamics of $V(t,w,x)$ can then be written as:

$$\max_{c,\omega}\left\{u(c,x) + V_t + V_w(r + \omega(\mu-r))w - V_w c + \frac{1}{2}(\sigma\omega w)^2 V_{ww} + V_x x g\right\} = 0.$$

See e.g. Rogers (2013) for a detailed derivation of this equation. Furthermore, due to the time-homogeneity of the payoff-function, one can easily check that $V(t,w,x) = e^{-\rho t} V(w,x)$, where $V(w,x)$ then must satisfy

$$\max_{c,\omega}\left\{u(c,x) + V_w(r + \omega(\mu-r))w - V_w c + \frac{1}{2}(\sigma\omega w)^2 V_{ww} + V_x x g - \rho V\right\} = 0. \quad (7)$$

Partial differentiation with respect to consumption $c$ and the equity share $\omega$ yields the first-order conditions

$$u_c(c^*, x) = V_w(w, x) \tag{8}$$

and

$$\omega^* = -\left(\frac{V_w}{V_{ww}w}\right)\left(\frac{\pi}{\sigma^2}\right), \tag{9}$$

respectively, where we recall that $\pi$ is the equity premium.

When (8) and (9) are substituted into the HJB equation, it becomes apparent that our model is scalable in the sense that the value function must share the homogeneity property of the utility function and thus be homogeneous of degree zero in $w$ and $x$. Furthermore, the policy function for consumption, defined implicitly by (8), can be proven to be homogeneous of degree one in $w$ and $x$, sharing this property with the consumption policy function of the classical Merton problem with a standard CRRA utility function. Indeed, combining (8) with the fact that $u_c$ is homogenuous of degree $-1$, it follows that

$$u_c\left(\frac{c^*(\alpha w, \alpha x)}{\alpha}, x\right) = V_w(w, x),$$

and therefore, we see that

$$c^*(\alpha w, \alpha x) = \alpha u_c^{-1}(V_w(w, x), x) = \alpha c^*(w, x).$$

By extension, the implied policy function for the consumption-wealth ratio is homogeneous of degree zero in the same two variables. In other words, the policy functions for the equity share and the consumption-wealth ratio depend on the wealth-norm ratio $w/x$ only.

Substituting the optimal values transforms the HJB equation into a partial differential equation (PDE) for the value function $V$ in the variables $w$ and $x$:

$$u(c^*(w,x),x) - \rho V(w,x) + V_w(w,x)\left[\left(r + \frac{1}{2}\omega^*(w,x)\pi\right)w - c^*(w,x)\right] + V_x(w,x)xg = 0$$

However, the homogeneity property that we just derived transforms it into an ordinary differential equation in $w$ only, for given fixed $x > 0$. That is, Euler's homogeneous function theorem implies that

$$V_x(w,x)x = -V_w(w,x)w.$$

Suppressing the arguments of $V$ and $V_w$, we can then write the HJB equation as

$$u(c^*,x) - \rho V + V_w\left[\left(r + \frac{1}{2}\omega^*\pi - g\right)w - c^*\right] = 0. \qquad (10)$$

The presence of the second-order derivative $V_{ww}$ in the expression (9) for the optimal equity share $\omega^*$ makes this a non-linear ODE of the second order in $w/x$.

We study this equation in more detail in the subsequent section, where certain analytical properties will be derived and studied. After this we will show that the Merton problem with our chosen utility function has an analytical solution, found through the dual value formulation.

### 2.3 Analytical properties

The form of (8) suggests that the policy function for consumption is continuous and monotonic on $(0,\infty)^2$. The explicit solution derived in the Appendix and discussed in the subsequent subsection confirms this suggestion. The monotonicity of the corresponding function for the equity share should be intuitive because an increase in wealth allows the agent to afford higher consumption; and because the curvature of the utility function decreases as consumption rises, we expect the equity share to rise. As seen from (9), this share is proportional to the reciprocal of the curvature of the value function. While the curvature of the utility function is a step function, the measure of curvature of the value function itself is continuous on $(0, \infty)^2$, again seen as a consequence of the analytical solution. A smooth graph of the equity-share policy function, based on the numerical solution presented in the next section, can be inspected in Figure 2 below.

Although the policy function for consumption is continuous everywhere, its derivative makes a jump at the point where $c^* = x$. To see this, differentiate first both sides of equation (8) with respect to $w$ and solve for the derivative of optimal consumption with respect to wealth. On elasticity form, it becomes

$$\frac{\partial \ln c^*}{\partial \ln w} = \frac{V_{ww} w}{u_{cc} c^*}.$$

Although $V_{ww}$ is continuous, $u_{cc}$ makes a jump at the point where consumption equals the norm. Let $w^x$ denote the wealth level at which this choice is optimal. We then see that the consumption policy function must have a kink at this wealth level. To explore this issue further, note first from (9) that

$$V_{ww} w = -\left(\frac{V_w}{\omega^*}\right)\left(\frac{\pi}{\sigma^2}\right).$$

Next, differentiating the utility function (4) twice, we find

$$u_c = -c^{-\gamma_j} x^{\gamma_j - 1}, \quad u_{cc} = -\gamma_j c^{-\gamma_j - 1} x^{\gamma_j - 1},$$

so that

$$u_{cc} c = -\gamma_j c^{-\gamma_j} x^{\gamma_j - 1} = -\gamma_j u_c,$$

where $j = 1,2$ depending whether $c < x$ or $c \geq x$, respectively. Furthermore, substituting for $\gamma_j$ for the optimal equity share for the preference ordering CRRA$_j$, as given in (2), we find

$$u_{cc} c = -\left(\frac{u_c}{\omega_j}\right)\left(\frac{\pi}{\sigma^2}\right).$$

Noting from (8) that $u_c = V_w$ in equilibrium, we the then find

$$\frac{V_{ww} w}{u_{cc} c} = \frac{\omega_j}{\omega^*}.$$

From the monotonicity of $\omega^*$, we thus have

$$\frac{\partial \ln[c^*(w,x)/w]}{\partial \ln w} = \begin{cases} \frac{\omega_1 - \omega^*}{\omega^*} < 0, w < w^x \\ \text{undefined}, w = w^x \\ \frac{\omega_2 - \omega^*}{\omega^*} > 0, w > w^x \end{cases} \qquad (11)$$

We lastly note some limiting properties. If, for given $x$, $w \to \infty$ the probability of optimal consumption falling below the norm becomes extremely low. In the limit, then, the agent will act as if the relevant preferences were CRRA$_2$, so that $\omega^* \to \omega_2$ and $c^*/w \to \eta_2$, where we recall that $\eta_2$ was defined in (3). The monotonicity of $\omega^*$ and the positive sign of (11) for $c^* > x$ furthermore warrant that both limits will be approached from below.

If, in contrast, $x \to \infty$ for given $w$, the agent will behave as if preferences were CRRA$_1$. Monotonicity implies that $\omega^*$ then will approach $\omega_1$ from above. Because its policy function is homogeneous in degree zero, this result suggests that the same will happen when $w \to 0$ for given $x$. The asymptotic properties furthermore indicate that the slope of this function will be rather flat on both ends, meaning that we expect the policy function for the equity share to resemble a sigmoid curve.

However, the negative sign of (11) for $c^* < x$ suggests that $c^*/w$ then will approach $\eta_1$ from below if $x \to \infty$ for given $w$. In practice, it makes less sense to consider cases of very high norms than very high wealth levels. Thus, for the parameter values that we will consider, we expect the optimal consumption-wealth ratio to start well below $\eta_1$ for low wealth levels and then dip even further with rising wealth levels until again rising with wealth when wealth levels are high enough to make optimal consumption higher than the norm.

## 2.4 Closed form solution to the dual value function.

Given a general utility function $u$ to be used in the infinite horizon Merton problem, one would not in general expect a closed form solution. However, our proposed utility function can be seen as a state dependent combination of two CRRA functions. It is well known that a closed form solution to the Merton problem can be found when the utility function is a simple CRRA function, and one is therefore tempted to think that these results should be readily extended to our case. By following for example Rogers (2013) Sec. 1.3, we see that the dual value function approach to solving the Merton problem will allow us to find certain closed form expressions, and we derive this method in full in the appendix but give here a brief overview of the findings.

As discussed earlier the utility function is divided into two regimes depending on the social norm $x$, namely through the indicators $\mathbb{1}_{w<x}$ and $\mathbb{1}_{w\geq x}$. We fix a level of the social norm at a value $x > 0$. By doing a change of variables, we will let $z$ implicitly represent the marginal utility of consumption, that is, we define

$$(t, z) = \big(t, V_w(t, w, x)\big),$$

where $(t, z) \in A := [0, \infty) \times \big(V_w(t, \infty, x), V_w(t, 0, x)\big)$. Define a new function $J(\cdot\,|x): A \to \mathbb{R}$ by $J(t, z|x) = V(t, w, x) - wz,$ and then through certain manipulations (explicitly proven in the Appendix) we find that the dual value function can be represented by

$$J(t, z|x) = F_1(t, z, x)(xz)^{1-1/\gamma_1} + F_2(t, z, x)(xz)^{1-1/\gamma_2} + G(t, z, x).$$

Here $F_1$ and $F_2$ are two functions capturing the weighted expectation of the fractional moment of a geometric Brownian motion conditioned on the event of being above or below the soft social norm $x$, and the function $G$ captures the weighted probabilities of the inverse of a geometric Brownian motion being above or below the norm $x$.

With this expression, we may now derive the quantities of interest based on $V(t, w, x)$, namely the consumption and equity share policy functions, by first observing that $J_z(t, z|x) = w$. Using this fact, differentiating the dual function and inverting, recalling that $z$ represents $V_w(t, w, x)$, one can derive an expression describing the marginal utility of consumption, and thus also derive an expression for the optimal consumption policy $c^*$. A similar procedure can be used to find the optimal equity share policy. We may illustrate this more simply in the classical CRRA utility case without state dependence, where the dual value function $J(t, z)$ would have a similar shape as above, i.e. $J(t, z) = z^{1-1/\gamma} h(t)$, for some function $h$. Due to the fact that $J_z(t, z) = w$, it is then readily checked that the marginal utility of consumption is given by $V_w(t, w) = w^{-\gamma} h(t)^\gamma$.

The representation obtained for the dual value function, and thus also the value function, is therefore a natural continuation of the solution to the classical Merton problem, but now taking into account the state dependent nature of our utility function, and the probabilities of ending up in these states.

For computational purposes, however, the functions $F_1$ and $F_2$ are rather complicated and involve an indefinite time-integral over certain cumulative probability functions with an argument depending on time. As illustrated above, in order to derive the quantities of interests, we need to invert the derivative of $J(t, z|x)$, which is certainly rather complicated, and would probably benefit from numerical methods. Still, the analytical representation of the dual value function provides valuable insight into the dynamics which is generated by our proposed utility function.

Alternatively, of course, one can solve the HJB equation derived in (10) directly by well-known computational methods, allowing for efficient and precise simulations for both the optimal portfolio weights and consumption. In the next section we will go deeper into these results.

## 3. Numerical Solution

We solved the ODE equation in (10) numerically under the assumption of a social norm of $x = 3$. Because of the model's homogeneity properties, the solution for one value of $x$ is sufficient for computing behavior at any other norm level. We chose this particular value mainly for ease of exposition. As shown in Table 1, we furthermore calibrated the utility function such that the standard curvature measure is $\gamma_2 = 2$ for above-norm consumption levels, but the much higher $\gamma_1 = 6$ for below-norm consumption. Our values of equity premium, stock-market volatility, and the subjective discount rate correspond to most of the literature. The risk-free rate of 2.5 percent may seem high compared to the experience of the 2010s. It seems more compatible with recent observations in the 2020s; furthermore, for comparison with habit models, it needs to be higher than the growth rate $g$ of the social norm. With these parameter values, the implied equity share is approximately 25 percent for CRRA$_1$ and 75 percent for CRRA$_2$.

*Table 1: Parameter values (annual rates)*

| Variable | Symbol | Value |
|---|---|---|
| Equity premium | $\pi$ | 4.8 pp |
| Equity return std. dev. | $\sigma$ | 17.89% |
| Riskless return (real) | $r$ | 2.5% |
| Subjective discount rate | $\rho$ | 4% |
| RRA for $c^* < h$ | $\gamma_1$ | 6 |
| RRA for $c^* \geq h$ | $\gamma_2$ | 2 |
| Social norm growth rate | $g$ | 1.9% |

Because our model is partial, it contains no productivity-driven natural growth rate that can be used to anchor the growth rate $g$ of the social norm. However, as optimal consumption in our model consistently stays below the expected portfolio return, wealth keeps rising in expectation, which makes consumption grow as well. We then calibrate the norm growth rate $g$ at 1.9 percent so as to roughly match the expected growth rate of consumption and wealth in the model[4].

Our solution method transforms the second-order ODE into a two-equation first-order ODE system, which we solve over a range $[w_{min}, w_{max}]$, where $w_{min} = 10$ and $w_{max} = 4,000$. For given guesses of the initial equity share and consumption-growth ratio, $\omega(w_{min}, x)$ and $\eta(w_{min}, x)$, we solved this system forward using the solution software for such problems in DifferentialEquations.jl, a package developed in the Julia programming language by Rackauckas and Nie (2017). Our initial guesses for $\omega(w_{min}, x)$ and $\eta(w_{min}, x)$ were derived from the corresponding values for the highly risk averse preference ordering CRRA$_1$. As our convergence criterion, we required the value of $V_w(w_{max}, x)$ to be close to the corresponding value for the preference ordering CRRA$_2$ with low risk aversion. By trial and error, we found the equity share for CRRA$_1$ of 25 percent to be a highly robust guess for the solution value $\omega(w_{min}, x)$, which allowed us to limit the search for initial guesses to one dimension, the consumption-wealth ratio. We were then able to solve the model as a saddle-point problem[5].

### 3.1. Policy Functions
Figure 2 displays the graph of the policy function for the equity share for wealth levels ranging from 10 to 2,000, which seems sufficient for characterizing behavior with $x = 3$. As just mentioned, the optimal equity share starts out as the same level as the one with the

---

[4] For $w \to \infty$, this growth rate in our model approaches $r + \omega_2 \pi - \eta_2$ independently of the parameter value for $g$. With our parametrization, this limiting value equals 1.95 percent.

[5] A combination of grid search and a golden-ratio search narrowed down the consumption-wealth ratio $\eta(w_{min}, x)$ to a very tight range. Unfortunately, the forward-looking instability at this point was so great that the precision needed to pin down the stable saddle-point solution exceeded the computer's precision level. However, all candidate solutions within the tight range were indistinguishable up to $w = 69.9$. Restarting the search from that point on allowed computation of a stable saddle-point solution satisfying the convergence criterion at $w = w_{max} = 4,000$. We then cut and pasted the two solutions together at $w = 69.9$ to obtain a complete solution.

*Figure 2: Policy function for the equity share*

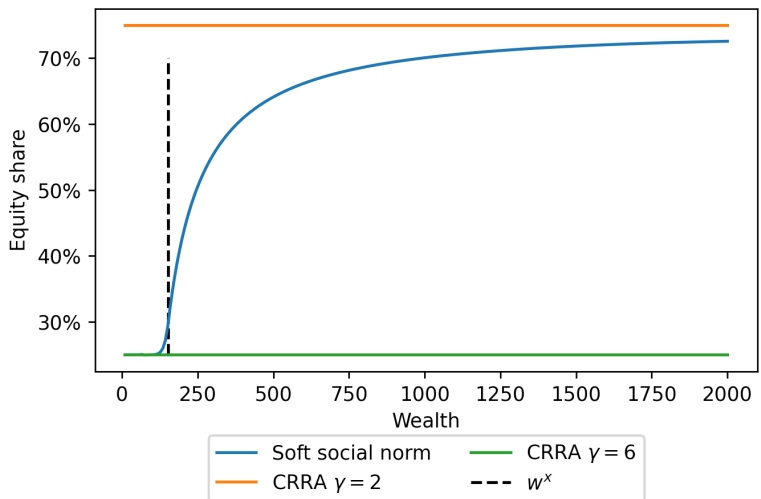

The red and green lines show the optimal equity shares for the CRRA preference orderings with relative risk aversion of 2 and 6, respectively. The dashed vertical line denotes the wealth level $w^x$ at which the optimal consumption equals the social norm.

highly risk averse preference ordering CRRA$_1$ at very low wealth levels. From there, it increases monotonically and continuously with the approximate shape of a sigmoidal curve and approaches the 75 percent level of the less risk averse preference ordering CRRA$_2$ when wealth is high enough for the probability to be very low of consumption falling below the norm. The rise from the bottom level starts at wealth levels somewhat below $w^x$, where optimal consumption just equals the social norm. It rises rather steeply around this level, reflecting the diminishing probability of having to consume less than the norm as wealth increases.

The procyclicality of risk taking is an important feature of our model. It may cause the agent to want to sell equity after prices drop because the loss of wealth makes the agent more risk averse. Similarly, the agent may want to "buy at the top" because risk aversion is lower in good times.

Figure 3 shows the policy function for consumption. For easier reading, this graph stops at $w = 500$. The green and red lines show, for comparison, the corresponding policy functions for CRRA$_1$ and CRRA$_2$, respectively. Whereas these are both linear, the policy function with the soft social norm consists of two slightly concave segments, joined together

*Figure 3: Policy function for consumption*

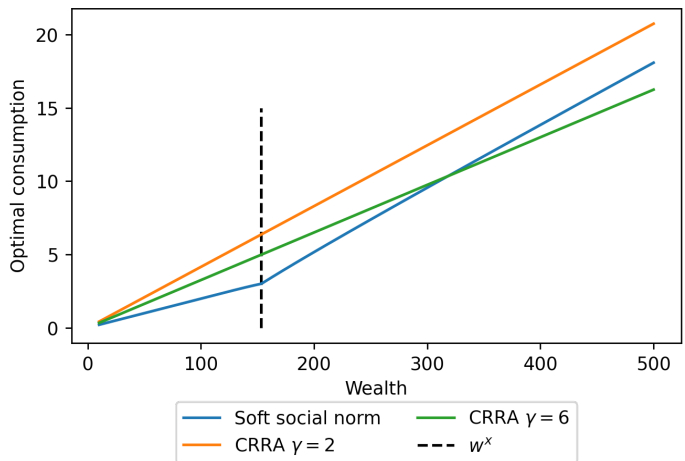

The red and green lines show optimal consumption for the CRRA preference orderings with relative risk aversion of 2 and 6, respectively. The dashed vertical line denotes the wealth level $w^x$ at which the optimal consumption equals the social norm.

with a kink at $w = w^x$. Although the graph rises everywhere, it is flatter when consumption falls below the social norm. In fact, as wealth rises towards this level, the slope becomes even a little bit flatter until suddenly becoming steeper at $w = w^x$. From there on, it is steeper than both alternatives, crosses the graph for CRRA$_1$, and ultimately approaches the one for CRRA$_2$, albeit at much higher wealth levels than the ones shown in this graph.

Taken together, Figures 2 and 3 reveal how our agent smooths consumption when wealth falls towards and/or below $w^x$. The main mechanism is that risk then is taken down, so that wealth fluctuates less. Some additional smoothing is obtained as the marginal propensity to consume (MPC) out of wealth falls abruptly when wealth drops below $w^x$. However, for wealth just above that level, the MPC is actually at its highest. So, when wealth falls to levels close to, but just above $w^x$, consumption is cut more than in proportion to the decline in wealth. The resulting rise in saving then helps avoid a further drop below $w^x$.

Figure 4 repeats the picture from Figure 3, but now in the form of the consumption-wealth ratio. Here, the optimal ratios for CRRA$_1$ and CRRA$_2$ become horizontal lines. The latter serves as an upper limit, toward which the policy function converges as wealth grows

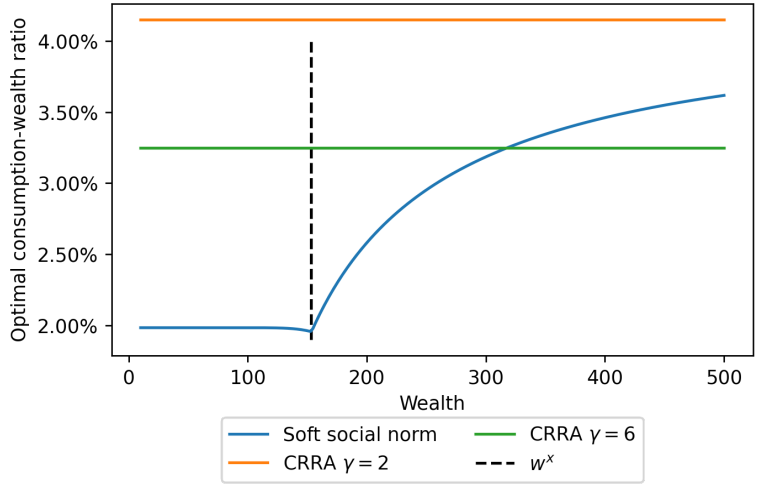

*Figure 4: Policy function for the consumption-wealth ratio*

The red and green lines show optimal consumption-wealth ratios for the CRRA preference orderings with relative risk aversion of 2 and 6, respectively. The dashed vertical line denotes the wealth level $w^x$. at which the optimal consumption equals the social norm.

very large. However, as noted in Section 2, the policy function starts well below the level that is optimal for CRRA$_1$ preferences and falls even a little bit further before optimal consumption reaches the social norm. Then, it rises sharply before levelling out gradually.

This is a case of dynamic substitution that seems rather unique to our specification. The motivation for consuming less than with CRRA$_1$ preferences at low wealth levels comes from the prospect of reaching beyond the norm if the agent can succeed in building sufficient wealth.

To see this more formally, consider the $s$-period marginal rate of intertemporal substitution

$$M_s = e^{-\rho s} \frac{u_c(c_{t+s}, x_{t+s})}{u_c(c_t, x_t)}$$

for pairs of consumption/norm ratios such that $c_{t+s}/x_{t+s} > c_t/x_t$. Let $M_s^1$ denote the corresponding factor for an individual with CRRA$_1$ preferences globally. Then, clearly, if $c_{t+s}/x_{t+s} < 1$, $M_s = M_s^1$. For $c_{t+s}/x_{t+s} \geq 1 > c_t/x_t$, it is easlily verified that $M_s = M_s^1 (c_{t+s}/x_{t+s})^{\gamma_1 - \gamma_2} \geq M_s^1$. And for $c_{t+s}/x_{t+s} > c_t/x_t \geq 1$,

$$M_s = M_s^1 \left(\frac{c_{t+s}/x_{t+s}}{c_t/x_t}\right)^{\gamma_1-\gamma_2} > M_s^1.$$

This means that our agent discounts the marginal utility of a brighter future less severely than the CRRA$_1$ agent and is thus willing to sacrifice more to potentially obtain it. This saving motive is reinforced by the expectation that the norm is expected to grow over time at rate $g$. The latter point became clear in experiments that we did with lower growth rates, which resulted in significantly higher consumption-wealth ratios at low wealth levels. With a faster growing social norm, it takes more saving to help consumption catch up, so that the incentive implicit in the stochastic discount factor applies to a greater amount of saving.

Figure 5 illustrates this dynamic substitution more clearly by displaying the optimal consumption-wealth ratio together with the expected portfolio return at each wealth level, where the latter, $r + \omega\pi$, varies with the equity share. The solid and dashed red lines show the corresponding consumption-wealth ratio and expected portfolio return, respectively, for the CRRA$_2$ preferences, which we think of as "no social norm." The distance between the expected portfolio return and the consumption-wealth ratio can be interpreted as an expected saving rate out of wealth. Starting from the lowest wealth level, the graph in Figure 5 then shows this saving rate rising with wealth as the consumption-wealth ratio falls even as the expected return rises with the equity share, reflecting the correspondingly improved chance of climbing out of below-norm consumption levels. After crossing the barrier $w^x$ it continues to rise. In fact, it rises above the corresponding rate for the case of no social norm even though the rate of portfolio return is much lower because of the low equity share. It then levels off and asymptotically approaches the saving rate of case with no social norm.

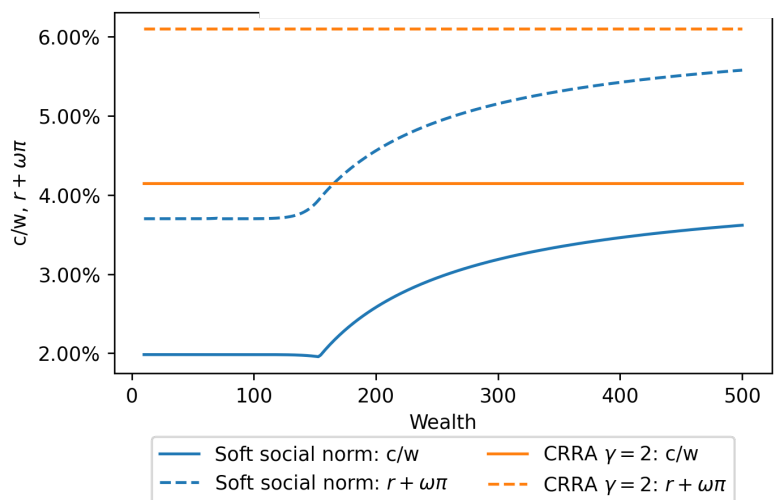

*Figure 5: Closeup of consumption-wealth ratio and expected return around $w^x$ revealing strong saving motive*

### 3.2. Behavior over time

We simulate the model solution over 50 years to study the dynamic development of wealth, risk taking, and consumption. The step size is specified as 12 per year, corresponding to months, so that the simulation covers 600 months. We draw 50,000 duplications and start with an initial wealth level implying consumption at 10 percent above the social norm. Figure 6 presents a fan chart for the distribution of future wealth relative to the social norm. It is increasingly skewed to the right, consistent with the well-known result that normally distributed instantaneous returns imply log-normal distributions for future wealth. However, the mean is pretty much horizontal, indicating a central tendency for consumption to grow at the same rate as the social norm. This result is not necessarily implied by the model, however. Although the consumption growth rate implied by the agent's choices depends on the growth rate of the social norm, this relationship is not one to one, at least not at all dates. However, as noted in Section 3, we have, by trial and error, calibrated the growth rate of the social norm so as to roughly match the average consumption growth.

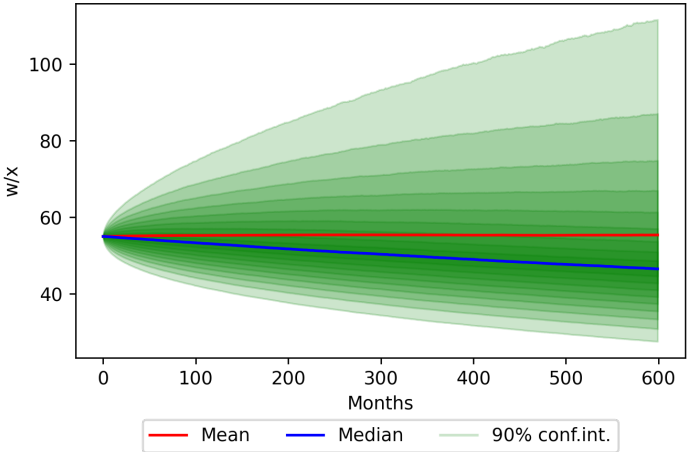

*Figure 6: Simulated development of wealth relative to the social norm*

Figure 7 presents a similar fan chart for the equity share, using the same draws. It is highly skewed to the right, which is natural given the lower limit given by $\omega_1$, which is about 25 percent. Even so, the mean is fairly stable across time horizons, which fits with our notion of consumption above the social norm as the normal case. However, because the norm grows over time, the equity share tends to remain well below than the 75 percent upper limit in our model.

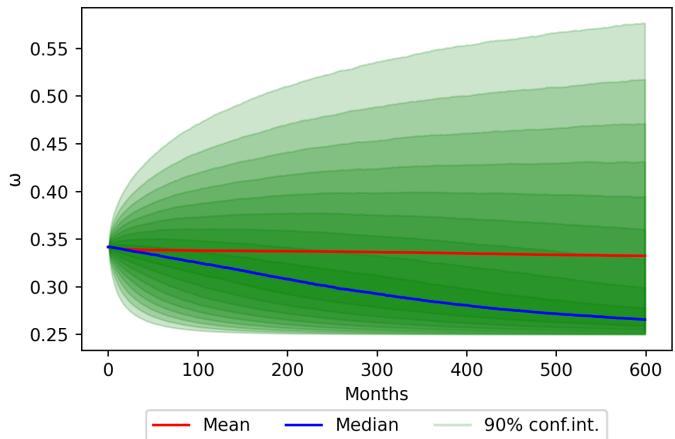

Figure 7: Simulated development of equity share

Figure 8 shows the corresponding pattern for consumption relative to norm. Because the MPC is lower for $w < w^x$ than for $w > w^x$ the skewness of the distributions is even more pronounced than for wealth. The kink at this point is reflected as a kind of corner indicated by an arrow. The lower MPC below this wealth level limits the distribution's downside, which makes the mean rise slightly with the horizon.

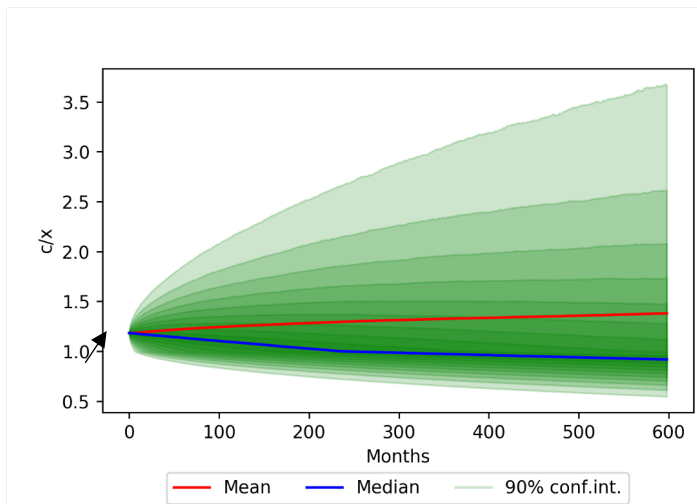

Figure 8: Simulated development of consumption relative to the social norm

Taken together, Figures 5, 7, and 8 illustrate the efforts that our agent undertakes to curtail the probability of having to cut consumption below the social norm. Although the initial wealth in our example is high enough to maintain the consumption-wealth ratio that a CRRA$_1$ agent would choose, our agent chooses to save more, thus making wealth grow faster than the social norm, which—in combination with low risk taking—puts a lid on the probability of financial outcomes that are sufficiently negative to call for below-norm consumption.

The incentive to stay away from the region where $c < x$ is naturally greater the higher the curvature of the utility function in that region. In the limiting case where $\gamma_1 \to \infty$, this incentive becomes absolute in the sense that utility falls to minus infinity in that region. For a solution to exist in this case, initial wealth must be large enough to safely support and sustain consumption at or above the ever-growing social norm, which requires

$$w_0 \geq \frac{x_0}{r - g} \qquad (12)$$

When this inequality binds, the agent will be constrained to always just consuming the social norm with all wealth invested in the riskless asset because otherwise the case of $c - x < 0$ will have a positive probability, which cannot be optimal[6]. Moving beyond this point will be impossible because the agent's income cannot increase when equity investment is kept at zero, which is necessary to avoid the risk of consumption falling below the norm. If the inequality is strict, however, the agent will be able to take some risk and raise saving and thus be able to maintain consumption higher than the social norm. The feasibility constraint in (12) resembles those found in most habit models, but with the important difference that, in our model, it applies only in the limiting case where $\gamma_1 \to \infty$.

---

[6] Clearly, $r > g$ is needed for (12) to be valid. This requirement is also needed in a typical habit model. Whether or not it is needed in our model depends on the values of the other parameters.

## 4. Discussion

This section discusses the implications of our model for the practice of portfolio management and spending policies of endowment funds and sovereign wealth funds. We start by comparing the implications of our model with the habit-formation and similar models in the literature.

### 4.1 Comparison with habit-formation models

We base this comparison on the instantaneous utility function (5). The policy functions for such an agent have been derived by Constantinides (1990) and were further explored by Lindset and Mork (2019). Such functions have an advantage over our model in that the decline in curvature as wealth grows there is continuous. However, because the curvature grows without limit as consumption falls towards the habit level, they function best if the habit can be understood as a fairly low fraction of actual consumption. Furthermore, and importantly, they imply an absolute requirement that consumption always be kept above the habit level. This requirement is satisfied endogenously because their agent in good times sets aside a sufficient part of wealth in the riskless asset to ensure sufficient funds to be available to defray the cost of maintaining the habit in bad times.

      However, an agent wanting to initiate behavior based on such preferences needs to have sufficient initial wealth to allow a large enough set-aside in safe investments to satisfy the initial habit. In our model, this requirement arises only in the limiting case of $\gamma_1 \to \infty$, as discussed above. Accordingly, the owners of an endowment fund or a sovereign wealth fund that consider switching to a strategy compatible with habit preferences will need to set

aside sufficient funds in safe assets to maintain what they consider habitual, i.e. minimal spending. As mentioned in the introduction, this requirement may conflict, sometimes seriously, with preconceived ideas about minimum spending. Indeed, it might very well cause problems if the Norwegian government, as owners of the GPFG, were to switch to a habit-constrained type of behavior at the present time. We furthermore find it intuitively appealing to allow consumption to fall below the social norm in bad times because the restraint on consumption in good times may seem unduly harsh. Thus, our specification allows for more flexibility both initially and over time.

Habit-formation models have been advanced as explanations of the equity premium puzzle (e.g. Constantinides, 1990, and Campbell and Cochrane, 1999) and the observed smoothness of aggregate consumption. Because our model is partial and not necessarily intended to describe the behavior of all agents in a general-equilibrium model, we make no claims regarding the equity premium puzzle.

Nor can we claim that our model offers a good explanation for the smoothness of aggregate consumption. It is true that consumption is smoothed slightly relative to wealth as illustrated by the decreasing consumption-wealth ratio for $w < w^x$ in Figures 3 and 4. However, the sharply rising ratio for $w > w^x$ pulls very much in the opposite direction. The standard deviation (in annual terms) of the log change in consumption in our model[7] is 9.1 percent for the parameter values in Table 1. That is lower than the 13.4 percent implied by the low-curvature specification CRRA$_2$, mainly because wealth fluctuates less for our more risk-averse agent. However, it is significantly higher than the 4.5 percent[8] implied by the high-curvature specification $CRRA_1$.

---

[7] The simulated results are averages of the time-series estimated standard deviations for each of the duplications.

### 4.2 Implications for endowment and sovereign wealth funds

The analysis of our specification is intended as a contribution to the ongoing discussion about portfolio and withdrawal rules for endowment funds and sovereign wealth funds such as the Norwegian GPFG. Because one of our concerns is that portfolio and withdrawal rules must be consistent with each other, we emphasize the implications of our analysis along both dimensions. Although all such funds typically have rules about withdrawals, these rules may be bypassed in practice when it comes to actual spending as short-term borrowing may be used to handle fluctuating spending needs. It may be worth pointing out that such deviations in fact have portfolio implications in that short-term borrowing is equivalent to a lowering of the riskless share of the portfolio. Moreover, the GPFG is, by statute, barred from making such bypasses in that the Government Pension Fund Act of 1990 prohibits loan-financed spending as long as there is money left in the fund[9]. The motivation for this rule is to prevent policy makers from using borrowing to circumvent the withdrawal rules in general. This legal detail makes the practical implementation of withdrawal rules so much more important for the GPFG.

---

[8] This number is almost exactly like the one obtained from quarterly U.S. data on per-capita consumption since 1947. That does not mean that our $CRRA_1$ specification explains U.S. aggregate data, however, because our model is very much partial whereas a test of the equity premium needs a general-equilibrium model. In particular, with a representative agent, it needs a model with zero net holdings of the risk-free asset, so that the equity share is constrained to unity. With CRRA preferences, that constraint would make the volatility of consumption equal to that of equity wealth, which we have calibrated at 17.9 percent.

[9] https://www.regjeringen.no/contentassets/9d68c55c272c41e99f0bf45d24397d8c/government-pension-fund-act-01.01.2020.pdf, accessed October 5, 2022. All borrowing done by the Norwegian government after the 1990 passing of this Act is for the purpose of funding policy financial institutions, the largest of which is the *Lånekassen* student loan agency.

The main GPFG withdrawal rule, referred to as the Fiscal Rule, passed by the Storting (Parliament) in 2001[10], is to allow payouts in support of the government's annual budget corresponding to the expected real return on the fund's balance at the beginning of the year, currently stipulated as 3 percent. If strictly adhered to, this rule would require spending to follow the ups and downs of financial markets, independently of perceived spending needs. However, the Fiscal Rule also includes provisions to allow for smoothing, which may be approximately represented as the MIT-Tobin rule[11]

$$c_t = \lambda \bar{\eta} w_t + (1-\lambda) c_{t-1}, \qquad (13)$$

where $\bar{\eta}$ is the target withdrawal rate (3 percent for the GPFG).

Our analysis implies a number of recommendations for changes in this rule. First, the target ratio $\bar{\eta}$, the expected real financial return, is much too high. It is not sustainable, as shown by Dybvig and Qin (2021). Although our model does not recommend a fixed target ratio, the range of recommended ratios displayed in Figure 4 is consistently lower than the one implied by the low-curvature preference ordering CRRA$_2$, which in turn is significantly lower than the expected rate of portfolio return.

Second, although our model implies some smoothing of payouts at wealth levels below the social norm, the pattern above this wealth level is very much the opposite. In particular, financial losses should call for spending cuts that are more than proportional to the financial loss. This result follows from a desire to save as a safeguard against the risk of future spending below the social norm. The message from our analysis is that the current rule does not take this risk sufficiently seriously. The widespread use of withdrawal rules based on smoothing rules such as the one in (13) may perhaps be explained by beliefs in mean-reverting equity prices. Cochrane's (2022) discussion may give some support to this belief. However, mean reversion also means that long-term rates of return will be lower than the short-term ones, as analyzed by Mork and Trønnes (2022).

---

[10] https://www.regjeringen.no/no/dokumenter/stmeld-nr-29-2000-2001-/id194346/, accessed November 2, 2022.

[11] The Norwegian Fical Rule is actually more complicated, on the one hand because the rule applies to the structural rather than the actual non-oil deficit and on the other hand because it allows for discretionary fiscal policy as temporary deviations from the rule. However, all such deviations must be funded directly from the GPFG. These details are further discussed and analyzed in Mork, Trønnes, and Bjerketvedt (2022).

Third, managing the risk of payouts falling below the social norm implies significant limitations on the share of risky assets in the portfolio. However, the equity share need not, and should not be constant. By being procyclical it should allow higher risk taking when wealth is high enough for the probability of falling below the social norm is modest, and vice versa. High financial returns should be followed by increases in the equity share and vice versa. That means that our agent sometimes may want to "buy at the top" and "sell at the bottom." Again, the motivation "at the bottom" is to limit the risk of falling and/or staying below the social norm. But similarly, because this risk is modest "at the top," it will then be prudent to harvest more of the equity premium.

### 4.3. Mandate and implementation

In principle, our investor could delegate all investment decisions to portfolio managers by communicating the policy functions of the optimal solution. In practice, simpler rules are preferred because they are easier to communicate, and they make monitoring of manager performance easier. This subsection offers an evaluation of the expected-utility cost of formulating mandates as simpler rules of thumb rather than the optimal solution.

We determine this cost by simulating each of the alternatives over 100 years[12] and comparing their implied expected utility, computed as the mean subjectively discounted utility, based on the soft-social-norm preferences. We use the same 50,000 duplications in all alternatives so as to avoid differences among alternatives being caused by sampling errors. We start with consumption at 10 percent above the social norm, so that consumption falling below the norm is not an immediate threat, but an event close enough to guard against. For each alternative, we then divide the difference in expected utility from the optimal case by marginal utility[13] so as to express the difference as a loss measured in consumption units. Lastly, by dividing this loss by the marginal value of the initial level of wealth, we express it as the equivalent to a percentage loss of initial wealth.

---

[12] Our reason for now extending the simulation to 100 years is that the discount factor for this horizon becomes very close to zero. With $\rho = 0.025$, $e^{-100\rho} = 0.018$.

As a first alternative, we consider a mandate where the equity share and the consumption-wealth ratio are held constant over time, but where the constant values are derived as optimal subject to this constraint. We determine the constrained-optimal constants as the pair of values that provide the highest simulated expected discounted utility over 100 years.

*Table 2: Comparison with rules of thumb*

Loss of expected utility expressed in units of pct loss of $w_0$

| | |
|---|---|
| Constrained optimal constant $\omega = 28.9\%$ and $c/w = 2.1\%$ | 0.3% |
| Constrained optimal values updated every 10 years | 0.1% |
| Optimal initial values held constant, then updated every 10 years | 1.8% |
| CRRA with constrained optimal $\omega$ | 21.2% |
| CRRA with constrained optimal $\omega$, smoothed, $\lambda = 0.5$ (annual) | 21.2% |

---

[13] For alternatives whose results are reasonably close to the unconstrained optimal solution, we use the derivative of the value function at the starting wealth level. For alternatives with significantly inferior results (in practice, the last two lines of Table 2), we use the average of the same derivative at the starting wealth level and the one yielding the same value of the value function as the expected discounted utility of that alternative.

The result is presented as the first line in Table 2. The estimated implied loss is small, in fact, so small that many would call it trivial[14]. It becomes even smaller if we assume that the values for equity share and the consumption-wealth ratio are updated by recomputing the same procedure for each simulated scenario every 10 years[15]. It remains less than two percent of the initial wealth if instead we assume that the fund owner requires managers to use values that are optimal on the initial date and, for reasons of convenience, keeps it constant for 10 years at a time.

These results, which are reported in the first three lines of Table 2, may seem astonishing considering the considerable range of optimal values displayed in Figures 2 and 4. The explanation is that the optimal solution chooses a combination of equity share and consumption-wealth ratio that makes consumption remain fairly stable relative to the social norm as the latter changes over time.

The constrained-optimal choice of equity share and consumption-wealth ratio thus depends on the starting point. If, instead of starting with consumption 10 percent above the norm, the agent starts with a wealth level implying consumption 10 percent *below* that norm, the constrained-optimal equity share is 27.2 percent and the consumption-wealth ratio 2.1 percent, both lower than with the higher initial wealth in Table 2. For much higher wealth levels, these values will naturally converge toward the optimal values for the unconstrained specification $CRRA_2$, which are 75.0 percent and 4.1 percent, respectively.

The near optimality of constant policy functions may give the impression that our preference ordering can be closely approximated by a CRRA function or Epstein-Zin preferences with constant risk parameter and elasticity of intertemporal substitution. That is not the case, however, as can be readily seen from the fact that the constrained-optimal combination of equity share $\omega = 28.9$ percent and consumption-wealth ratio of $c/w = 2.1$ percent could not have been implied by any of the two alternatives. For expected CRRA preferences and our values for equity variability and equity premium, an equity share of 28.9 percent would have required a curvature parameter of $\gamma = 5.2$, which in turn would have implied a consumption-wealth ratio as high as $c/w = 3.3$ percent. A higher curvature parameter would have implied a lower consumption-wealth ratio, but also a lower equity share. In fact, for expected CRRA, the consumption-wealth ratio cannot fall below the risk-free rate, which in our case equals 2.5 percent. With Epstein-Zin preferences, the corresponding lower bound is defined as the smaller of the riskless rate and the subjective discount rate. With the latter calibrated at 4 percent, the lower bound is again 2.5 percent.

---

[14] For a fund worth USD 1.3 trillion, 0.3 percent is still USD 3.9 billion. We leave it to the reader to judge whether that is trivial.

[15] Updates with approximate 10-year intervals have been the practice so far of the GPFG.

Thus, with our parameter values, the constrained-optimal values for the consumption-wealth ratio and the equity share lie outside the possible combinations for expected CRRA utility and Epstein-Zin preferences. Although this may not be the case for other sets of parameter values or other starting points for wealth, our example clearly shows that the constrained-optimal policy functions for soft social norms can be quite different from these frequently used preference specifications. This is another example of the essential role that dynamic substitution plays in our specification. High saving helps build wealth despite low equity income and thus serves as a relative safeguard against suffering consumption below the social norm.

We considered including an alternative based on the rule of equating consumption to the expected financial return, with or without smoothing, and with varying degrees of risk taking. Unfortunately, such rules fit poorly into our framework because we have ignored non-financial income[16] that would typically grow over time to match the trend growth of the social norm. Because consumption, when set to equal the expected financial return, would then not grow in expectation, there would be a steady tendency for consumption to fall behind the exogenously determined social norm over time. As a result, we would exaggerate the cost of following such rules because they look worse in our partial framework than they would do in a more realistic, but also more complicated framework with non-financial income.

As a substitute, we consider an alternative based on the behavior of a person with CRRA expected-utility preferences with risk aversion implying the same equity share as in the constrained-optimal case of the first line in Table 3. With our parameter values, that requires a curvature parameter of $\gamma = 5.2$ which, as noted above, would imply a consumption-wealth ratio of 3.3 percent, about 50 percent higher than in the constrained optimum. The welfare loss from following that behavioral pattern, shown in the fourth line of Table 2, corresponds to a more than 20 percent loss of initial wealth, two orders of magnitude greater than the constrained optimum.

As a final alternative, we consider a spending rule like the one in (13), but where $\bar{\eta}$ is the consumption-wealth ratio of the CRRA preferences with $\gamma = 5.2$, and $\lambda = 0.056$ per month, implying an annual rate of adjustment[17] of about 0.5. The result is shown in the last line of Table 2.

Clearly, the smoothing does not help. In fact, it makes the result slightly worse, albeit only in the second decimal of the percentage, which is not shown. The benefits of smoothing are small as long as consumption stays above the social norm; and smoothing has the negative effect of widening the distribution of future wealth levels. The latter effect is almost negligible in our case, however, making the net effect of smoothing essentially nil.

---

[16] For individual investors, this would typically be labor income. For endowment funds it would be current revenue such as tuition payments, and for sovereign wealth funds current tax revenues.
[17] 0.056 is the approximate solution to the equation $(1 - \lambda_m)^{12} = 1 - \lambda_a$, where $\lambda_a = 0.5$.

## 5. Summary and Conclusions

Inspired by debates about investment and withdrawal strategies for endowment funds and sovereign wealth funds, this paper has explored the implications a CRRA preference ordering modified by a soft social norm as a possible representation of investor preferences. Although we find this representation appealing, our purpose has not necessarily been to promote it as normatively better than alternatives, but to explore how the behavior implied by this model differs from that of other models. Even so, we believe this exploration could be useful in terms of guiding the concrete decisions about risk taking and withdrawal policy that will have to be made by investors whose preferences are influenced by factors similar to those we consider.

Our model shares the two leading features of habit models and other models with lower bounds on consumption, namely, smoother consumption than with CRRA preferences and high and procyclical risk aversion, albeit lower risk aversion than the typical habit model. The main difference is that our agent has a much higher propensity to save in bad times. Our agent is quite willing to substitute consumption dynamically by saving a lot when wealth is low in the hope of building sufficient wealth to allow consumption above the social norm in the future. Models of habits and other lower bounds on consumption do not capture this effect because there consumption above the bound is secured by sufficient holding of safe assets, which in turn requires a severe constraint on consumption in good times. By allowing higher consumption in good times as well as somewhat higher risk taking in bad times, we believe our model may more closely resemble the preferences revealed by the actions observed for large endowment funds and sovereign wealth funds.

That said, our analysis suggests a number of changes that the owners of such funds may want to implement if they share our belief in the importance of social spending norms. Annual spending should be significantly lower than expected financial returns. Spending as a share of wealth should furthermore be procyclical; in particular, financial losses should, as a rule, be followed by larger than proportional spending cuts so as to build enough new wealth to limit the probability of consumption falling below the norm. The only exception would be to keep spending from falling too far below the social norm.

The desire to keep consumption above the social norm carries important implications for financial risk taking. It should be modest, although not as modest as in the typical habit-formation models. It should be procyclical, again not to the same degree as with habit formation, but enough so that the fund's managers sometimes may want to "buy at the top" and "sell at the bottom."

Because the policy functions implied by our model are highly non-linear, delegating them to professional managers may not be simple. However, at least with our parameter values and other assumptions, we find that the welfare loss would be quite modest if mandates for equity share and withdrawals are updated only every 10 years, even if the mandates are set myopically at the values that would have been currently optimal if updates had been made continuously. However, this mandate would depend on the wealth owner's initial wealth and look very differently from one based on expected CRRA utility or Epstein-Zin preferences.

The exogeneity of the social norm may be considered a limitation of our model. A promising modification might be to replace this soft norm by an equally soft habit, where the habit is built up from past consumption rather than exogenously. The resulting model would still differ quite a bit from the habit-formation literature by treating the habit as a soft rather than a hard bound on consumption. We intend to turn to this task in a follow-up paper.

## Appendix: Analytical solutions to the Merton problem through the dual value formulation.

While the proposed utility in (4) behaves very differently than the standard CRRA utility function, we are able to derive an analytic solution by using the dual value function approach to solve the HJB equation associated to the value function.

To this end, we will follow the procedure from Rogers (2013) Sec. 1.3. First, it is completely clear from the time-independence of the utility function in (4) that it satisfies a standard transversality condition, i.e. that $\lim_{t \to \infty} e^{-\rho t} u_c(c, x) = 0$. The dual value function approach is based on a generalized Legendre transform, and for this it is convenient with a change of variables. Define the new coordinates

$$(t, z) = (t, V_w(t, w))$$

where $(t, z) \in A := [0, \infty) \times (V_w(t, \infty), V_w(t, 0))$. Define a new function $J: A \to \mathbb{R}$ by

$$J(t, z) = V(t, w) - wz.$$

Note that the variable $z$ now implicitly describes the optimal control, since by first order conditions the marginal utility of the optimal consumption $c^*$ is equal to $V_w(t, w)$. We define $\tilde{u}(z, x)$ to be the convex dual of $u(c, x)$, that is

$$\tilde{u}(z, x) = \sup_c \{u(c, x) - zc\}.$$

This is a decreasing function in $z \mapsto \tilde{u}(z, x)$, and it is readily seen that $\tilde{u}(z, x) = u((u_c)^{-1}(z, x), x) - z(u_c)^{-1}(z, x)$. It is readily checked that the inverse of the $(u_c)^{-1}(z, x)$ is given by

$$(u_c)^{-1}(z, x) = \mathbb{1}_{z^{-1/\gamma_1} x^{1-1/\gamma_1} \leq x} z^{-1/\gamma_1} x^{1-1/\gamma_1} + \mathbb{1}_{z^{-1/\gamma_2} x^{1-1/\gamma_2} > x} z^{-1/\gamma_2} x^{1-1/\gamma_2}.$$

Inserting the inverse function into $u$ we can compute that

$$u((u_c)^{-1}(z, x), x)$$

$$= \mathbb{1}_{z^{-1/\gamma_1} x^{1-1/\gamma_1} \leq x} \frac{1}{1 - \gamma_1} \left[ (zx)^{1-1/\gamma_1} - 1 \right]$$

$$+ \mathbb{1}_{z^{-1/\gamma_2} x^{1-1/\gamma_2} > x} \frac{1}{1 - \gamma_2} \left[ (zx)^{1-1/\gamma_2} - 1 \right].$$

And so using this representation and subtracting $z(u')^{-1}(z, x)$ we find that

$$\tilde{u}(z, x) = \mathbb{1}_{z^{-1} \leq x} (1 - 1/\gamma_1)^{-1} \left[ (zx)^{1-1/\gamma_1} - 1 \right] + \mathbb{1}_{z^{-1} > x} (1 - 1/\gamma_2)^{-1} \left[ (zx)^{1-1/\gamma_2} - 1 \right].$$

By time homogeneity of the payoff function we know that $V(t, w) = e^{-\rho t} v(w)$, and therefore, the convex dual satisfies $J(t, z) = e^{-\rho t} j(ze^{\rho t})$. From Sec. 1.3 in Rogers (2013) it then follows that the convex dual is given by the following conditional expectation

$$e^{\rho t} J(t, z) = E\left[ \int_0^\infty e^{-\rho s} \tilde{u}(Y_s, x) \, ds \Big| Y_0 = e^{\rho t} z \right].$$

Here, $Y$ is a geometric Brownian motion satisfying

$$dY_t = Y_t \left( \left| \frac{\mu - r}{\sigma} \right| dW_t + (\rho - r) dt \right).$$

Observe that by the Fubini theorem, we can take expectations inside the integral, and we are left to compute

$$E[\tilde{u}(Y_s, x) | Y_0 = e^{\rho t} z]$$

$$= E\left[ \mathbb{1}_{Y_s^{-1} \leq x} (1 - 1/\gamma_1)^{-1} \left[ (Y_s x)^{1-1/\gamma_1} - 1 \right] \Big| Y_0 = e^{\rho t} z \right]$$

$$+ E\left[ \mathbb{1}_{Y_s^{-1} > x} (1 - 1/\gamma_2)^{-1} \left[ (Y_s x)^{1-1/\gamma_2} - 1 \right] \Big| Y_0 = e^{\rho t} z \right]$$

$$= J^1(s, t, z, x) + J^2(s, t, z, x)$$

Manipulating further, using the explicit solution to the geometric Brownian motion, we see that

$$J^1(s,t,z,x) = e^{(1-1/\gamma_1)\rho t}(1-1/\gamma_1)^{-1}(zx)^{1-1/\gamma_1} E\left[\mathbb{1}_{Y_s^{-1} \leq x}(Y_s)^{1-1/\gamma_1}\big|Y_0 = 1\right]$$

$$- (1-1/\gamma_1)^{-1} E\left[\mathbb{1}_{Y_s^{-1} \leq x}\big|Y_0 = e^{\rho t}z\right]$$

$$= C_1(t)(zx)^{1-1/\gamma_1} N_1(s,x) - (1-1/\gamma_1)^{-1} P(Y_s^{-1} \leq x|Y_0 = e^{\rho t}z).$$

Here, $N_1(s,x)$ represents a conditional expectation of a power of a geometric Brownian motion, and $P(y < x|A)$ is the conditional probability that the Log-normally distributed variable $y$ is less than $x$ given $A$. A similar representation can be made for $J^2$. Our goal is now to find an expression for the conditional expectations $N_i$ for $i = 1,2$. By Itô's formula it follows that for $i = 1,2$, $Z_t^i = (Y_t)^{1-1/\gamma_i}$ is a geometric Brownian motion with mean $(1-1/\gamma_i)\left(\rho - r - \frac{1}{\gamma_i}\left|\frac{\mu-r}{\sigma}\right|^2\right)t$ and standard deviation $(1-1/\gamma_i)\left|\frac{\mu-r}{\sigma}\right|\sqrt{t}$. We therefore need to evaluate two conditional expectations of two log-normally distributed random variables $\sim Log\mathcal{N}\left((1-1/\gamma_i)\left(\rho - r - \frac{1}{\gamma_i}\left|\frac{\mu-r}{\sigma}\right|^2\right)t, (1-1/\gamma_i)^2\left|\frac{\mu-r}{\sigma}\right|^2 t\right).$

This can be computed explicitly using the following formula: suppose $Y \sim Log\mathcal{N}(\bar{\mu}, \overline{\sigma^2})$, and let $\Theta_{\bar{\mu},\overline{\sigma^2}}$ denote the cumulative probability distribution associated to a normally distributed variable $\sim \mathcal{N}(\bar{\mu}, \overline{\sigma^2})$. Then for any $h > 0$ we have that

$$E[Y|Y > h] = e^{\bar{\mu}+\frac{\overline{\sigma^2}}{2}} \frac{1 - \Theta_{\bar{\mu}+\overline{\sigma^2},\overline{\sigma^2}}(ln(h))}{1 - \Theta_{\bar{\mu},\overline{\sigma^2}}(ln(h))}$$

and

$$E[Y|Y < h] = e^{\bar{\mu}+\frac{\overline{\sigma^2}}{2}} \frac{\Theta_{\bar{\mu}+\overline{\sigma^2},\overline{\sigma^2}}(ln(h)) - \Theta_{\bar{\mu},\overline{\sigma^2}}(ln(h))}{1 - \Theta_{\bar{\mu},\overline{\sigma^2}}(ln(h))}.$$

Summarizing our findings, we see that the dual value function $J(t,z,x)$ can be written as

$$J(t,z|x) = F_1(t,z,x)(xz)^{1-1/\gamma_1} + F_2(t,z,x)(xz)^{1-1/\gamma_2} + G(t,z,x),$$

where $F_1$ and $F_2$ are two functions capturing the weighted conditional expectation of the fractional moment of a geometric Brownian motion, and the function $G$ captures the weighted probabilities of the inverse of a geometric Brownian motion being above or below the log of the soft social norm. In particular, one can readily check that $z \mapsto F_i(t,z,x)$ for $i = 1,2$ and $z \mapsto G(t,z,x)$ is smooth on $[0,\infty) \times (0,\infty)^2$. To now find the value function, one may follow the procedure outlined in Section 2.5.

## References:

Abel, Andrew. B., 1990. "Asset Prices under HabitFormation and Catching up with the Joneses." *American Economic Review,* 80 (2, Papers and Proceedings): 38-42.

Alhashel, B., 2015. "Sovereign Wealth funds: A literature review" *Journal of Economics and Business,* Volume 78: 1-13.

Arouri, Mohamed, Sabri Boubaker, Wafik Grais, and Jocelyn Grira. 2018. "Rationality or politics? The color of black gold money. " *The Quarterly Review of Economics and Finance* 70: 62–76.

Arrow, Kenneth, Partha Dasgupta, Lawrence Goulder, Gretchen Daily, Paul Ehrlich, Geoffrey Heal, Simon Levin, Kar-Göran Mäler, Stephen Schneider, David Starret, and Brian Walker, 2004. "Are We Consuming Too Much?" *Journal of Economic Perspectives*, 18 (3): 147-172.

Arun, Thillaisundaram, 2012. "The Merton Problem with a Drawdown Constraint on Consumption." arXiv:1210.5205.

Baldwin, C., 2012. *Sovereign Wealth funds: The New Intersection of Money and Politics.* Oxford: Oxford University Press.

Barber, B. M. & Wang, G., 2013. "Do (some) university endowments earn alpha?" *Financial Analysts Journal,* 69(5), pp. 26-44.

Bernstein, Shai, Josh Lerner, and Antoinett Schoar, 2013. "The Investment Strategies of Sovereign Wealth Funds." *Journal of Economic Perspectives* 27: 219–37.

Braunstein, Juergen. 2022. *Capital Choices: Sectoral Politics and the Variation of Sovereign Wealth*. Ann Arbor: University of Michigan Press.

Brown, J. R., Dimmock, S. G., Kang, J.-K. & Weisbrenner, S. J., 2014. "How University Endowments Respond to Financial Market Shocks: Evidence and Implications." *American Economic Review,* Volume 104, pp. 931-962.

Campbell, J. & Deaton, A., 1989. "Why is Consumption So Smooth?" *Review of Economic Studies,* Volume 56, pp. 357-374.

Campbell, J. Y. & Cochrane, J. H., 1999. "By Force of Habit: A Consumption-Based Explanation of Aggregate Stock Market Behavior." *Journal of Political Economy,* 107(2), pp. 205-251.

Campbell, John Y. and Ian W. R. Martin, 2022. "Sustainability in a Risky World." Harvard University and London School of Economics.

Campbell, John Y. and Roman Sigalov, 2022. "Portfolio choice with sustainable spending: A model of reaching for yield." *Journal of Financial Economics*, 143: 188-206.

Cochrane, J. H., 2022. "Porfolios for Long-Term Investors." *Review of Finance*: 1-42.

Constantinides, G. M., 1990. "Habit Formation: A Resolution of the Equity Premium Puzzle." *Journal of Political Economy,* 98(3), pp. 519-543.

Dahiya, S. & Yermack, D., 2018. "Investment returns and distribution policies of non-profit endowment funds". Working Paper 25323, National Bureau of Economic Research.

Dreassi, A., Miani, S. & Paltrinieri, A., 2017. "Sovereign Pension and Social Security Reserve Funds: A Portfolio analysis. " *Global Financial Journal,* Volume 34, pp. 43-53.

Duesenberry, James S. (1949). *Income, Saving, and the Theory of Consumer Behavior*. Cambridge, Mass., Harvard University Press.

Dybvig, Philip H., 1995. "Duesenberry's Racheting of Consumption: Optimal Dynamic Consumption and Investment Given Intolerance for any Decline in Standard of Living." *Review of Economic Studies*, 62 (2): 287-313.

Dybvig, Philip H., 1999. "Using Asset Allocation to Protect Spending." *Financial Analyst Journal*: 55 (1): 49-62.

Galí, J., 1990. "Finite horizons, life-cycle savings, and time-series evidence on consumption" *Journal of Monetarh Economics,* Volume 26, pp. 433-452.

Irarrazabal, Alfonso A., Lin Ma, and Juan Carlos Parra-Alvarez, 2020. "Optimal Asset Allocation for Commodity Sovereign Wealth Funds." BI Norwegian Business School.

Jeon, Junkee and Kyunghyun Park, 2021. "Portfolio selection with drawdown constraint on consumption: a generalization model." *Mathematical Methods of Operations Research* 93: 243-289.

Johan, S., Knill, A. & Mauck, N., 2013. "Determinants of Sovereign Wealth Fund Investment in Private Equity vs. Public Equity." *Journal of International Business Studies,* Volume 44, pp. 155-172.

Lindset, S. & Mork, K. A., 2019. "Risk Taking and Fiscal Smoothing with Sovereign Wealth Funds in Advanced Economies." *International Journal of Financial Studies,* 7(4), pp. 1-24.

Merton, R. C., 1971. "Optimum consumption and portfolio rules in a contnuous-time model." *Journal of Economic Theory* 3 (4): 373-413.

Mork, Knut Anton and Haakon Andreas Trønnes, 2022. "Expected long-term rates of return when short-term returns are serially correlated." Norwegian University of Science and Technology (NTNU).

Mork, Knut Anton, Haakon Andreas Trønnes, and Vegard Skonseng Bjerketvedt, 2022. "Capital Preservation and Current Spending with Sovereign Wealth Funds and Endowment Funds: A simulation Study." *International Journal of Financial Studies* 10: 1-24.

Paltrinieri, A. & Pichler, F., 2013. "Asset Management Issues in Sovereign Wealth Funds: An Empirical Analysis." In: J. Falzon, ed. *Bank Performance, Risk and Securitization.* London: Palgrave Macmillan.

Rackauckas, C. & Nie, Q., 2017. "DifferentialEquations.jl-2017 – A Performant and Feature-Rich Ecosystem for Solving Differential Equations in Julia." *The Journal of Open Research Software,* Volume 5.

Rogers, L. C. G., 2013. *Optimal Investment*. London: Springer.

Shin, Yong Hyun, Byung Hwa Lim, and U. Jin Choi, 2007. "Optimal consumption and portfolio selection problem with downside consumption constraints." *Applied Mathematics and Computation* 188: 1801-1811.

Tobin, J., 1974. "What is permanent endowment income?" *American Economic Review,* 64(2), pp. 427-432.

van den Bremer, Ton, Frederick van der Ploeg, and Samuel Willis. 2016. "The Elephant in the Ground: Managing Oil and Sovereign Wealth." *European Economic Review* 82: 113–31.

World Commission on Environment and Development, 1987. *Our Common Future*. United Nations.